\documentstyle[12pt]{article}
\newcommand{\beq}{\begin{equation}}
\newcommand{\eeq}{\end{equation}}
\newcommand{\bea}{\begin{eqnarray}}
\newcommand{\eea}{\end{eqnarray}}

\begin{document}
\thispagestyle{empty}
\begin{flushright}
NTUTH-98-066 \\
June 1998
\end{flushright}
\bigskip\bigskip\bigskip
\vskip 2.5truecm
\begin{center}
{\LARGE {Chiral structure of the solutions of}}
\vskip 7pt
{\LARGE {the Ginsparg-Wilson relation}}
\end{center}
\vskip 1.0truecm
\centerline{Ting-Wai Chiu\footnote{E-mail address: twchiu@phys.ntu.edu.tw}, 
Chih-Wei Wang, Sergei V. Zenkin\footnote{Permanent 
address: Institute for Nuclear
Research of the Russian Academy of Sciences, 117312 Moscow, Russia.
E-mail address: zenkin@al20.inr.troitsk.ru}}
\vskip5mm
\centerline{Department of Physics, National Taiwan University} 
\centerline{Taipei, Taiwan 106, R.O.C}
\vskip 2cm
\bigskip \nopagebreak \begin{abstract}
\noindent

We analyse the structure of solutions of the Ginsparg-Wilson relation for 
lattice Dirac operator in topologically trivial gauge sector. We show that 
the properties of such solutions relating to the perturbative stability
of the pole of the fermion propagator as well as to the structure of the 
Yukawa models based on these solutions are solely determined by the non-local 
chirally invariant part of these Dirac operators. Depending on the structure 
of this part, the pole in the fermion propagator may or may not be stable 
under radiative corrections. We illustrate this by explicit examples.\\


\end{abstract}
\vskip 1.5cm

\newpage\setcounter{page}1

\section{Introduction}
 
The Ginsparg-Wilson (GW) relation \cite{GW} for lattice Dirac operator
opens very interesting new possibilities for formulation of vector gauge 
theories, in particular QCD, on a lattice \cite{Hea}--\cite{Cha}. In its 
simplest form the GW relation reads as
\beq
D \gamma_5 + \gamma_5 D = 2 r D \gamma_5 D,
\label{GW1}
\eeq
where $D$ is the lattice Dirac operator and $r$ is a nonzero real parameter
(the lattice spacing $a$ is set to one).

It is commonly regarded that equation (\ref{GW1}) by itself ensures two
crucial, and usually mutually exclusive, properties of the operator $D$:
the absence of additive mass renormalization and reproducing correct anomaly, 
in particular the Atiyah-Singer index theorem.
However, as shown in \cite{CZ}, only the GW relation itself is not sufficient
to guarantee that any $ D $ satisfying it reproduces the
index theorem on the lattice. A necessary condition for the
solutions $D$ of eq.\ (\ref{GW1}) to have nonzero index in topologically
nontrivial sectors has been formulated in \cite{CZ}.
In topologically trivial sector the general
solution of eq.\ (\ref{GW1}) is given in terms of the chirally invariant
Dirac operator $D_c$ \cite{CZ}:
\beq
D = \frac{D_c}{1+rD_c}, \quad D_c \gamma_5 + \gamma_5 D_c = 0.
\label{sol2}
\eeq
In this paper we demonstrate that although the form (\ref{sol2})
forbids the additive mass corrections in the
fermion propagator determined by $D$, it does not ensure the perturbative
stability of the pole of the propagator. As we show, the question 
whether the pole
is stable under radiative corrections or not is determined solely by the
structure of chirally invariant operator $D_c$. We illustrate this by explicit
examples. We also show that in the absence of gauge interactions
the Yukawa models formulated by L\"uscher \cite{Lu} with the GW-Dirac
operators $D$ are equivalent to the ordinary Yukawa models defined with the 
corresponding operators $D_c$. 

In section 2, we discuss the basic properties of $D_c$ and
consider three explicit examples for which the operator $D$ for free 
fermions is local in the continuum limit. In section 3 we examine the 
fermion self-energy determined by these operators, and in section 
4 discuss the Yukawa models. Section 5 is a summary.

\section{Explicit solutions of the Ginsparg-Wilson relation}

For $D$ is nonsingular, according to eq.\ (\ref{sol2}), the problem of
constructing explicit solutions of $D$ reduces to finding a proper
realization of the chirally invariant Dirac operator $D_c$.
We are interested in those $ D $ which are local at least in the continuum
limit, i.e. $ D_{m n} $ vanishes at $a \rightarrow 0$ for any finite
physical distance $a|m-n|$. This implies that at
large $|m-n|$, $D_{m n}$ should fall off faster than $|m-n|^{-1}$.
On the regular lattice this means that for free fermions 
the Fourier transform of $D_{m n}$,
$D(p)$, should be a continuous periodic function of momentum $p$,
and the smoother the function,
the faster $D_{m n}$ falls off with $|m-n|$. This imposes certain 
limitations to the form of $D_c$: $D_c(p)$ should be either continuous, or 
may have some singularities, but should not have finite discontinuities.

According to the Nielsen-Ninomiya theorem
\cite{NiN} any continuous $D_c(p)$ suffers from species doubling, so 
$ D(p) $ constructed via
eq.\ (\ref{sol2}) with such $D_c$ will suffer from the doubling 
too\footnote{The simplest example of such $D$ is that constructed 
with the naive Dirac operator $D_c(p) = i \gamma_{\mu} \overline{p}_{\mu}$.
This example clearly demonstrates that the GW relation is not a remedy 
for the species doubling.}.
To avoid species doubling in $D_c$, and therefore in $D$, $D_c(p)$ should be 
discontinuous, i.e. $ D_{c} $ should be nonlocal in position space.
Such operators can be classified according to the
behaviour of $D_c(p)$ at the boundary of the Brillouin zone, where
$D_c(p)$ has either finite or infinite discontinuities. In order to have
$D(p)$ continuous, one should reject the $D_c(p)$ with finite
discontinuities, like the operators involving SLAC derivative \cite{SLAC}.
At present we know only three examples of explicit $ D_c $
whose Fourier transform in the free fermion limit has infinite 
discontinuities. \\

\noindent (I) The first example is the action proposed by Rebbi in \cite{Re}:
\beq
D_c(p) = i \gamma_{\mu} \overline{p}_{\mu} \: \frac{\hat{p}^2}
{\overline{p}^2}, \quad \overline{p}_{\mu}=\sin p_{\mu},
\quad \hat{p}_{\mu}
= 2 \sin \frac{p_{\mu}}{2}.
\label{Rep}
\eeq
$D_c(p)$ is singular at those points at the boundary of the Brillouine zone 
where the naive Dirac operator $i \gamma_{\mu} \overline{p}_{\mu}$ has zeroes 
which produce species doubling. In the presence of gauge field, $D_c$ takes 
the form
\beq
D_c = 4\left\{\gamma_{\mu} (\nabla^{+}_{\mu} 
+ \nabla^{-}_{\mu}),
(-\nabla^{-}_{\nu}\nabla^{+}_{\nu} + c g \sigma_{\nu \sigma} 
F_{\nu \sigma})^{-1}\right\}^{-1},
\label{Ren}
\eeq
where 
\bea
&&\nabla^{+}_{\mu \: m n} = U_{m, m+\hat{\mu}} \delta_{m+\hat{\mu},n}
- \delta_{m,n} \cr
&&\nabla^{-}_{\mu \: m n} = \delta_{m,n} - U_{m, m-\hat{\mu}} 
\delta_{m-\hat{\mu},n},
\label{nab}
\eea
$U_{m, m\pm\hat{\mu}}=\exp[\pm ig A_{\mu}(n\pm\hat{\mu}/2)]$,
$\sigma_{\mu \nu} = i [\gamma_{\mu}, \gamma_{\nu}]/2$, $F_{\mu \nu}$ is a 
lattice transcription of the 
field tensor, $c$ is a constant which in \cite{Re} has been chosen equal to
$1/2$, and $\{A,B\}=AB+BA$. \\  
 
\noindent (II) The second example is the Dirac operator which follows 
from the construction of the fermion path integral for the Weyl ordering 
\cite{Ze}:
\beq
D_c(p) = i \gamma_{\mu} \tilde{p}_{\mu}, \quad \tilde{p}_{\mu} 
= 2 \tan \frac{p_{\mu}}{2}.
\label{Wep}
\eeq
In this case $D_c(p)$ is singular at the whole Brillouin boundary.
In the position space it can be written in the form
\beq
D_c = \sum_{\mu} \gamma_{\mu} \left\{ (\nabla^{+}_{\mu} 
+ \nabla^{-}_{\mu}), (4 + \nabla^{+}_{\mu} 
- \nabla^{-}_{\mu})^{-1} \right\}.
\label{Wen}
\eeq

\noindent (III) Our third example is the chirally invariant operator
$D_c$ deduced from the overlap-Dirac operator proposed by Neuberger in 
\cite{Ne}. This $D$ satisfies 
eq.\ (\ref{GW1}) for $r \in (1/4, \infty)$ and therefore has
the form of eq.\ (\ref{sol2}) with
\bea
D_c(p) = i \gamma_{\mu} \overline{p}_{\mu} \: \frac{2M}{\overline{p}^2}
\left[\frac{1}{2}\hat{p}^2 - M + \sqrt{\overline{p}^2
+\left(\frac{1}{2}\hat{p}^2 - M\right)^2}\right],
\label{Nep}
\eea
where $M \in (0, 2)$ is intrinsic parameter of the formulation
related to $r$ as $M = (2r)^{-1}$. $D_c$ in (\ref{Nep})
has singularities at the same points of the Brillouin boundary
as Rebbi's operator (\ref{Rep}).
In the presence of gauge interactions, $D_c$ can be written as
\beq
D_c = 2M \frac{1+V}{1-V}, \quad
V = D_W \frac{1}{\sqrt{D_{W}^{\dagger} D_{W}}},
\label{NeD_c}
\eeq
where
$D_{W}$ is the Wilson operator with negative mass $-M$,
\beq
D_W = \frac{1}{2} [\gamma_{\mu} (\nabla^{+}_{\mu} 
+ \nabla^{-}_{\mu})
-\nabla^{-}_{\nu}\nabla^{+}_{\nu}] - M.
\label{DW}
\eeq
In terms of the unitary operator $V$, eq.\ (\ref{sol2}) takes the form
\beq
D = M(1+V).
\label{NeD}
\eeq

\section{Fermionic self energy}
\subsection{General structure}
When $D$ is nonsingular, eq.\ (\ref{sol2}) is equivalent to
\beq
D^{-1} = D_c^{-1} + r.
\label{sol1}
\eeq
Then the fermion propagator in the presence of dynamical gauge fields is
\beq
\langle D^{-1} \rangle = \langle D_c^{-1} \rangle + r,
\label{pro}
\eeq
where the brackets denote normalized averaging over all topologically trivial
gauge configurations. In order to simplify the analysis of self-energy 
operators, we limit ourselves to the quenched approximation,
i.e.\ $\mbox{det}D = 1$ in the measure of averaging in 
(\ref{pro}). Then in the momentum space the full propagators $\langle 
D^{-1} \rangle$ and $\langle D_c^{-1} \rangle$ can be written as
\beq
\langle D^{-1} \rangle(p) = \frac{1}{D(p)+\Sigma_{GW}(p)}, \quad 
\langle D_c^{-1} \rangle(p) = \frac{1}{D_c(p)+\Sigma_c(p)}, 
\label{Sig}
\eeq
where $\Sigma_{GW}(p)$ and $\Sigma_{c}(p)$ are the fermion self-energy
operators related to the Dirac operators $ D $ and $ D_c $, respectively.
Since $D(p)$ and $D_c(p)$ vanish at $p=0$, it follows from (\ref{pro}) and 
(\ref{Sig}) that at $p \rightarrow 0$
\beq
\Sigma_{GW}(p) \rightarrow \frac{\Sigma_c(p)}{1+r\Sigma_c(p)}.
\label{SDD_c1}
\eeq

Operator $D_c$ is chirally invariant, so is $\Sigma_c$. This implies that it
contains only odd combinations of $\gamma$-matrices.
In the simplest case one has
\bea
&&\Sigma_c = i \gamma^{\mu} \Sigma_{c}^{\mu}, \cr
&&\Sigma_{GW}(0) = \frac{i \gamma^{\mu} \Sigma_{c}^{\mu}(0)}
{1+r^2 \Sigma_{c}^{2}(0)} + \frac{r \Sigma_{c}^{2}(0)}{1+r^2 \Sigma_{c}^{2}
(0)}
\label{SDD_c2}
\eea

If $\Sigma_c(0) = 0$, the poles in the propagators are not shifted,
and radiative corrections lead only to a renormalization of the fermion
wavefunctions. However, because of the nonlocality of $D_c$,
$\Sigma_c(0) $ may not be always zero.
Then the poles of both propagators are shifted, and in $\langle D^{-1}
\rangle$ a scalar additive term 
\beq
\delta m = \frac{r \Sigma_{c}^{2}(0)}{1+r^2 \Sigma_{c}^{2}(0)}
\label{dm}
\eeq
is generated.

In the perturbation theory $\Sigma = \sum_n g^{2n} \Sigma^{(2n)}$, and 
in the lowest order one has  
\beq
\Sigma_{GW}^{(2)}(p \rightarrow 0) \rightarrow \Sigma_{c}^{(2)}(p 
\rightarrow 0).
\label{SDD_c3}
\eeq
So there is no $\delta m$ generated in the lowest order in the perturbation
theory, however the pole still may be shifted.

\subsection{$\Sigma(0)$: one loop analysis}
 
In the perturbation theory one has 
\beq
D = D^{(0)} + g D^{(1)} + g^2 D^{(2)} + O(g^4),
\label{Dp}
\eeq
where in the momentum space $D^{(1,2)}$ can be written in the form
\bea
&&D^{(1)}(p',p) = \int_q \delta_{2\pi}(p'-p-q)V_{\mu}(p,q) A_{\mu}(q), \cr
&&D^{(2)}(p',p) = \int_{q',q} \delta_{2\pi}(p'-p-q'-q)
V_{\mu\nu}(p,q',q) A_{\mu}(q')A_{\nu}(q),
\label{ver}
\eea
with $\int_q = \int_{(-\pi/a, \pi/a]^4} d^4q/(2\pi)^4$ and $\delta_{2\pi}(q)$ 
is $2\pi$-periodic $\delta$-function. Then
\bea
\label{Si2}
\Sigma^{(2)}(p)&=&C_2 \int_q [-V_{\mu}(p+q,-q)\frac{1}{D^{(0)}
(p+q)} V_{\nu}(p,q) + V_{\mu\nu}(p,-q,q)] \Delta_{\mu \nu}(q) \cr
&=&\Sigma_1(p) + \Sigma_2(p), 
\eea
where $C_2$ is the Casimir operator, and $\Delta_{\mu \nu}$ comes from the 
gauge field propagator; in the covariant gauge it has the form 
\beq
\Delta_{\mu \nu}(q) = 
\delta_{\mu\nu}\frac{1}{\hat{q}^2}-(1-\alpha)\hat{q}_{\mu} \hat{q}_{\nu}
\frac{1}{(\hat{q}^2)^2}.
\label{del}
\eeq
In (\ref{Si2}) the first term in the
integrand is the lattice counterpart of the
continuum self-energy diagram, while the second term corresponds to the 
tadpole diagram, and they are denoted by $\Sigma_1$ and $\Sigma_2$,
respectively.

>From the Ward identities for the vertices
\bea
&& V_{\mu}(p,q)\hat{q}_{\mu} = D^{(0)}(p+q)-D^{(0)}(p), \cr
&& V_{\mu\nu}(p,q',q)\hat{q}'_{\mu}\hat{q}_{\nu} \cr
&& \quad = \frac{1}{2}[D^{(0)}(p+q'+q)
   -D^{(0)}(p+q')-D^{(0)}(p+q)+D^{(0)}(p)],
\label{Wi}
\eea
it follows that 
$\Sigma^{(2)}(p \rightarrow 0)$ does not depend on gauge fixing parameter 
$\alpha$ in (\ref{del}), so we fix the Feynman gauge $\alpha = 1$. 

The explicit form of the vertices for each of our examples are derived from 
the corresponding expressions for the operator $D$ through the expansion 
(\ref{Dp}).
By virtue of relation (\ref{SDD_c3}) one can instead of $D$ use only its 
chirally invariant part $D_c$. Except for the
operator in eq.\ (\ref{Wen}), the explicit expressions for the vertices are 
quite cumbersome, so we present here only the results for 
$\Sigma^{(2)}(p \rightarrow 0)$.\\

\noindent (I) $ D_c $ in eq.\ (\ref{Ren}): In this case we can use the results
obtained in \cite{BoK}.
Then we get
\beq
\Sigma^{(2)}_{c}(p \rightarrow 0) \rightarrow i \gamma_{\mu} 
p_{\mu} \frac{1}{p^2}
(A-c^2B),
\label{ReS}
\eeq
where $A$ and $B$ are certain finite momentum integrals. 
$B$ is determined by the $c$-part in eq.\ (\ref{Ren}) and depends on the 
lattice transcription of the field tensor $F_{\mu \nu}$.
It is evident that we cannot obtain $\Sigma_{c}^{(2)}(p \rightarrow 0) = 0$
without fine-tuning the parameter $c$. Thus, this example demonstrates 
perturbative instability of the pole of the fermion propagator. \\

\noindent (II) $ D_c $ in eq.\ (\ref{Wen}): $\Sigma^{(2)}_{c}(0) = 0$, both
${\Sigma_c}_1(0)$ and ${\Sigma_c}_2(0)$ vanish. \\  

\noindent (III) $ D $ in eq.\ (\ref{NeD}):
$\Sigma_{GW}^{(2)}(0) = 0$\footnote{This result is anticipated in view of
previous results obtained in the domain wall and the overlap formulations 
\cite{dwm}.}. In this case it is 
the result of cancellation of nonzero $ {\Sigma_{GW}}_1(0) $ and 
$ {\Sigma_{GW}}_2(0)$:
\bea
&&{\Sigma_{GW}}_1(0)=- {\Sigma_{GW}}_2(0) \cr
&& \quad =-C_2\int_q \frac{M}{\left(M+\sqrt{\overline{p}^2+B^2(p)}\right)^2}
\sum_{\mu}\left\{V^{2}_{\mu}(0,q)[1-\overline{B}(q)]\right.\cr
&&\quad\quad\quad+\left.W^{2}_{\mu}(0,q)[1+\overline{B}(q)]
+2V_{\mu}(0,q)W_{\mu}(0,q)\overline{C}_{\mu}(q)\right\}
\frac{1}{\hat{q}^2},
\label{SN}
\eea
where
\bea
&&B(q)=\frac{1}{2}\hat{p}^2 - M, \quad \overline{B}(q)
=\frac{B(q)}{\sqrt{\overline{p}^2+B^2(p)}}, \cr
&&\overline{C}_{\mu}(q)=\frac{\overline{p}_{\mu}}
{\sqrt{\overline{p}^2+B^2(p)}}, \cr            
&&V_{\mu}(p,q)=\cos(p_{\mu}+\frac{1}{2}q_{\mu}), \quad
W_{\mu}(p,q)=\sin(p_{\mu}+\frac{1}{2}q_{\mu}).
\eea

\subsection{$\Sigma(0)$: mean field estimate}

Let us estimate now $\Sigma(0)$ in a mean-field approximation in which
$\langle U_{m, m\pm\hat{\mu}} \rangle 
=\langle U \rangle
= \mbox{const.} < 1$ 
and $\langle F(U) \rangle=F(\langle U \rangle)$ for any matrix $F(U)$. 
This approximation 
corresponds to the summation of certain subsets of the tadpole diagrams
in the Feynman gauge.
We have no illusions about its accuracy, however we find it quite 
instructive. For our examples we have: \\

\noindent (I) $\Sigma_{c}(p \rightarrow 0)$ has the same form as the r.h.s. 
of eq.\ (\ref{ReS}) with $A=2(1/\langle U \rangle -1)$, and 
$B=0$. From eq.\ (\ref{dm}) it follows that $\delta m = 1/r$ for any 
$\langle U \rangle < 1$. \\

\noindent (II) $\Sigma_{c}(0)=0$, $\delta m = 0$ for any 
$\langle U \rangle < 1$. \\

\noindent (III) $\Sigma_{GW}(0)=0$, $\delta m = 0$, if $\langle U \rangle >
1-M/4$. Note that this result agrees  
with that obtained in the domain wall fermions in \cite{ANZ}, as well as 
with the reasonings of ref.\ \cite{Ne2}.

\section{Yukawa models}

General solutions of the GW relation in terms of the chirally invariant
operator $D_c$ given in (\ref{sol2}) suggests a simple interpretation of 
the chirally and flavour invariant construction of Yukawa models proposed
in \cite{Lu}. In this secton we turn off the gauge interactions.

The fermion action in this formulation is defined as
\beq
S = \sum_n[\overline{\psi} D \psi -r^{-1} \overline{\chi} \chi
+y(\overline{\psi}+\overline{\chi})\Phi(\psi+\chi)],
\label{LY}
\eeq
where $\Phi = \frac{1}{2}(1+\gamma_5)\phi^{\dagger}+\frac{1}{2}(1-\gamma_5)
\phi$ is the Higgs field, $y$ is the Yukawa coupling, and 
$\overline{\chi}$, $\chi$ are auxiliary non-propagating fermion fields.
This action is invariant under the following infinitesimal variations
\bea
&&\delta \psi = \alpha \gamma_5 [(1-rD) \psi + \chi], \quad
\delta \chi = \alpha \gamma_5 rD \psi, \cr
&&\delta \overline{\psi} = [\overline{\psi} (1-rD) 
+ \overline{\chi}] \alpha \gamma_5, \quad 
\delta \overline{\chi} = \overline{\psi} rD \alpha \gamma_5, \cr
&&\delta \Phi = -\{\alpha \gamma_5, \Phi\}.
\eea
where $ \alpha $ is an infinitesimal global parameter which may be nontrivial
in the flavour space.
Making obvious change of variables and taking
into account eq.\ (\ref{sol2}) we obtain
\bea
&& Z[\Phi] = \int\prod_n d\psi d\overline{\psi} d\chi d\overline{\chi} \exp(-S)
= C \int\prod_n d\psi' d\overline{\psi}' \exp(-S') \cr
&& S' = \sum_n \overline{\psi}' (D_c +y \Phi) \psi'
\label{D_cY}
\eea
where $C$ is irrelevant constant, and the action $S'$ is chirally
invariant in the usual sense, i.e.\ invariant under the variations
\beq
\delta \psi' = \alpha \gamma_5 \psi', \quad
\delta \overline{\psi}' = \overline{\psi}'\alpha \gamma_5, \quad 
\delta \Phi = -\{\alpha \gamma_5, \Phi\}.
\eeq
Thus, the Yukawa model (\ref{LY}) with 
GW-Dirac operator $D$ is equivalent to the usual Yukawa model with the 
chirally invariant Dirac operator $D_c$.

System (\ref{D_cY}) can be easily examined within the mean-field approximation
developed in \cite{ToZ}. In particular, in the case of
$\phi^{\dagger}\phi = 1$ and scalar hopping parameter $\kappa \geq 0$ its
critical lines in this approximation are determined by only two constants
$G^W = \int_p [D_c^{\dagger}(p)D_c(p)]^{-1}$ and
$G^S = \int_p D_c^{\dagger}(p)D_c(p)$ (for more detail see \cite{ToZ}).

\section{Summary}

We have demonstrated that in topologically trivial sector some important
properties of a Dirac operator $D$ satisfying the GW relation (\ref{GW1})
are determined by the non-local chirally invariant Dirac operator $D_c$
introduced in (\ref{sol2}). In particular, in the absence
of the gauge interactions the Yukawa model constructed by L\"uscher \cite{Lu}
is shown to be equivalent to the ordinary Yukawa model defined with the 
operator $D_c$. In addition to the overlap-Dirac operator \cite{Ne}, we
constructed two new explicit solutions of the GW relation and examined the 
fermionic self-energy operators of these three operators respectively.
We demonstrate that a lattice Dirac operator $ D $ satisfying the GW relation
does not necessarily provide the perturbative stability of the pole of the
fermion propagator, and this property solely depends on the structure of the
corresponding chirally invariant Dirac operator $D_c$. Among the
three examples we have considered only two 
[ eqs.\ (\ref{Wen}) and (\ref{NeD_c} ] maintain
perturbative stability of the pole of the fermion propagator,
while the operator eq.\ (\ref{Ren}) fails without fine tuning the
parameter $c$, see eq.\ (\ref{ReS}).
 
On the other hand, in the topologically non-trivial sectors, the necessary
condition for a Dirac operator $D$ satisfying GW relation to have non-zero
index is $\mbox{det}(1 - r D ) = 0 $, that is equivalent to the non-existence 
of the chirally invariant operator $D_c$ in topologically nontrivial 
gauge field background \cite{CZ}. This condition is satisfied only by the
$D_c$ of the overlap-Dirac operator (\ref{NeD_c})--(\ref{NeD})
(for more detail see \cite{Chiu}, \cite{CZ}). We also note that
in two-dimensional gauge models operators $D_c$ in (\ref{Ren})
and (\ref{Wen}) have no genuine zeromodes in topologically non-trivial
sectors, similar to that of the naive Dirac operator \cite{CZ}.

\bigskip

This work is supported by the National Science Council, R.O.C. under
the grant number NSC87-2112-M002-013. \


\begin{thebibliography}{99}

\bibitem{GW} P.~H.~Ginsparg and K.~G.~Wilson, Phys.\ Rev.\ D 25 (1982) 2649.
%
\bibitem{Hea} P.~Hasenfratz, Nucl.\ Phys.\ B (Proc.\ Suppl.) 63A-C (1998) 53;
P.~Hasenfratz, V.~Laliena and F.~Niedermayer, ``The Index
Theorem in QCD with a finite cutoff", hep-lat/9801021;
P.~Hasenfratz, ``Lattice QCD without tuning, mixing and
current renormalization", hep-lat/9802007.
%
\bibitem{Ne} H.~Neuberger, Phys.\ Lett.\ B 417 (1998) 141; 
``More about exactly massless quarks on the lattice", hep-lat/9801031.
%
\bibitem{Lu} M.~L\"uscher, ``Exact chiral symmetry on the lattice and
the Ginsparg-Wilson relation", hep-lat/9802011.
%
\bibitem{Chiu} T.~W.~Chiu, ``Topological Charge and the Spectrum of
Exactly Massless Fermion on the Lattice", hep-lat/9804016,
to be published in Phys. Rev. D.
%
\bibitem{Cha} S.~Chandrasekharan, ``Lattice QCD with Ginsparg-Wilson
fermions", hep-lat/9805015.
%
\bibitem{CZ} T.~W.~Chiu and S.~V.~Zenkin, ``On solutions of the 
Ginsparg-Wilson relation'', hep-lat/9806019.
%
\bibitem{NiN} H.~B.~Nielsen and M.~Ninomiya, Nucl. Phys. B185
(1981) 20 [E: B195 (1982) 541]; ibid. B193 (1981) 173.
%
\bibitem{SLAC} S.~D.~Drell, M.~Weinstein and S.~Yankielowicz, Phys.
Rev. D 14 (1996) 1627.
%
\bibitem{Re} C.~Rebbi, Phys. Lett. B 186 (1987) 200.
%
\bibitem{Ze} S.~V.~Zenkin, Sov. Phys. Lebedev Inst. Rep. 9 (1988) 10;
Mod. Phys. Lett. A 6 (1991) 151.
%
\bibitem{BoK} G.~T.~Bodwin and E.~V.~Kovacs, Phys. Lett. B 193 (1987) 283.
%
\bibitem{dwm} S.~Aoki and Y.~Taniguchi, ``One-loop calculation in lattice QCD
with domain wall quarks'', hep-lat/9711004; H.~Neuberger, Y.~Kikukawa and 
A.~Yamada, ``Exponential suppression of radiatively induced mass in the 
trancated overlap'', hep-lat/9712022.
%
\bibitem{ANZ} S.~Aoki, K.~Nagai and S.~V.~Zenkin, Nucl. Phys. B (Proc. Suppl.)
63A-C (1998) 602.
%
\bibitem{Ne2} H.~Neuberger, Phys. Rev. D 57 (1998) 5417.
%
\bibitem{ToZ} S.~Tominaga and S.~V.~Zenkin, Phys. Rev. D 50 (1994) 2604;
S.~V.~Zenkin, Phys. Rev. D 53 (1996) 6674.

\end{thebibliography}
\end{document}